\documentclass[aps,prc,tightenlines,floats,floatfix]{revtex4}

\def\bea{\begin{eqnarray}}
\def\eea{\end{eqnarray}}

\def\be{\begin{equation}}
\def\ee{\end{equation}}
\def\ba{\begin{eqnarray}}
\def\ea{\end{eqnarray}}

\usepackage{graphics}
\usepackage{graphicx}
\usepackage{epsf} 
\usepackage{amsmath}
\usepackage{amssymb}

\begin{document} 

\phantom{0}
\vspace{-0.2in}
\hspace{5.5in}
\parbox{1.5in}{ \leftline{JLAB-THY-08-857}}

\vspace{-1in}

\title
{\bf Electromagnetic form factors of the $\Delta$ in a S-wave approach}
\author{G. Ramalho$^{1,2}$ and M.T. Pe\~na$^{2,3}$ 
\vspace{-0.1in}  }

\affiliation{
$^1$Thomas Jefferson National Accelerator Facility, Newport News, 
VA 23606 \vspace{-0.15in}}
\affiliation{
$^2$Centro de F{\'\i}sica Te\'orica e de Part{\'\i}culas, 
Av.\ Rovisco Pais, 1049-001 Lisboa, Portugal \vspace{-0.15in}}
\affiliation{
$^3$Department of Physics, Instituto Superior T\'ecnico, 
Av.\ Rovisco Pais, 1049-001 Lisboa, Portugal}

\vspace{0.2in}
\date{\today}

\phantom{0}

\begin{abstract} 
Without any further
adjusting of parameters, a relativistic constituent quark model,
successful in the description of the data
for the nucleon elastic form factors and of the dominant contribution for the 
nucleon to  $\Delta$ electromagnetic transition, is used here to
predict the dominant electromagnetic
form factors of the $\Delta$ baryon.
The model is 
based on a simple $\Delta$ wave function corresponding to
a quark-diquark system in an S-state.
The results for E0 and M1 are consistent both with experimental
results and lattice calculations.
The remaining form factors E2 and M3 vanish, given 
the symmetric structure taken for the $\Delta$.
\end{abstract}


\vspace*{0.9in}  
\maketitle

\section{Introduction}
\label{secI}

The internal quark structure of the nucleon can explain the experimental results for its electromagnetic form factors.
For instance, in Ref. \cite{Nucleon,Gross08b},  we presented a successful description
of the most recent results from Jlab for
the proton elastic form factors \cite{JlabData}.
Also, in the last few years there has been
an increasing convergence between the
lattice QCD extrapolations and 
the low momentum  ($Q^2 < 2$ GeV$^2$) form factor data 
\cite{Gockeler05,Wang07}.
In principle the $\Delta$ structure could also be inferred from the 
$\gamma N \to \Delta$ electromagnetic
transition \cite{Gockeler05,Wang07} (throughout this paper we will use ${\rm N}\Delta$ for short to name this reaction).
However, models based on quark degrees of freedom alone fail 
in the description of the dominant multipole 
(magnetic dipole $G_M^\ast$) 
of the
${\rm N}\Delta$ transition
\cite{NDelta,Newnucleondelta}.
Only phenomenological
models including also explicit pion
degrees of freedom, by adding coupled baryon-meson channels
to the constituent quark structure, can describe this transition \cite{SatoLee,Kamalov}.
Nevertheless, there are
uncertainties because there is no unique way of separating the pion cloud from 
the short distance quark core effects. 
\cite{Meissner07}.

As for lattice calculations, at present they are not yet successful 
in describing the N$\Delta$ 
electromagnetic form factors. Linear
extrapolations to the physical limit of
quenched calculations
underestimate the data for low $Q^2$ and
overestimate the experimental data for larger $Q^2$ \cite{Alexandrou05}.
The available unquenched lattice QCD results are
not sufficient for an extrapolation to the physical region
\cite{Alexandrou07}.
On the other hand, extrapolations of the lattice results
using Chiral Perturbation Theory ($\chi$PT)
were until now restricted to very low $Q^2$ \cite{Pascalutsa05,Gail06}.
Additionally, no $Q^2$ dependence of the form factors was studied 
using $\chi$PT. It is therefore important to understand
the structure of the $\Delta$,  in particular 
the quark distribution inside the $\Delta$ system, 
and to parametrize its electromagnetic form factors 
as functions of $Q^2$.

However, due to the short $\Delta$ mean life time, 
there is almost no experimental information 
about the $\Delta$ form factors, to the exception of 
the electrical charge.  The 
magnetic dipole form factor was measured at $Q^2=0$ only, giving
the magnetic moment $\mu_\Delta$ of the $\Delta^{++}$, since 
$\mu_\Delta= G_{M1}(0) \frac{e}{2M_\Delta}$
(where $M_\Delta$ is the $\Delta$ mass and $e$ the electron charge).
Our knowledge of the $\Delta^{++}$ dipole magnetic moment 
is essentially based on the experiments of Refs.~\cite{SIN,UCLA}.
The results of the analysis are not all
totally consistent \cite{Exp} nevertheless.
The Particle Data Group \cite{Yao06} gives a very broad 
interval for the $\Delta^{++}$ magnetic moment, $[3.7,7.5] \mu_N$, 
where $\mu_N $ is the nucleon  magneton.
The recent analysis using 
a Dynamical Model \cite{Castro01} 
quotes $\mu_\Delta=(6.14 \pm 0.51) \mu_N$, 
but the theoretical error is not included. 
As for the $\Delta^+$, the only available measurement is from Mainz
\cite{Kotulla02},
$\mu_\Delta = \left(2.7^{+3.5}_{-3.6}\right) \mu_N$, which still has 
a large error bar, mostly due to the theoretical 
uncertainties.
However, new experiments are planned, 
and methods based on a Chiral Effective-field theory \cite{Pascalutsa05b}
and Dynamical Models \cite{Chiang05} were  
proposed to extract the magnetic moment of  $\Delta^+$ 
from the reaction $\gamma p \to \pi^0 p \gamma^\prime$.
There is no data related to the $\Delta^0$ and $\Delta^-$ 
(see Refs. \cite{Castro01,Kotulla07} for a review).

In order to understand the present
disagreement between the
lattice calculations and the
experimental results for the electromagnetic
N$\Delta$ transition, 
theoretical constituent quark
models are  
promising tools. 
Given the absence of experimental data
for the $\Delta$,
they are needed to guide lattice extrapolations. Inversely
it is crucial to probe models away from the physical limit by
lattice data.

Models allow the calculation of the quark distribution 
inside the $\Delta$ and generate predictions for the form factors.
For the electrical quadrupole (E2)  there
are calculations by Buchmann
based on several models \cite{Buchmann}.
For the magnetic moments there is also a variety of 
predictions based on several frameworks, 
such as the SU(6) quark model \cite{Chiang05},
relativistic quark models \cite{Schlumpf93}, 
quark models with two-body exchange currents
\cite{Buchmann01} and sea quark contributions \cite{Riska06},
QCD sum rules \cite{Lee98,Aliev00},
Chiral and Soliton models 
\cite{Butler94,Kim,Lorce08,Hashimoto08}, 
field theoretical formulations with hadron 
degrees of freedom \cite{Machavariani08} 
and quark degrees of freedom \cite{Machavariani08b},   
and 
extrapolations of lattice QCD 
\cite{Pascalutsa05,Leinweber92,Cloet03,Lee05,Alexandrou07b}. 
We present a summary of these results in one
of the following sections (Table \ref{tableSummary}).
Almost all works are focused on the $\Delta$ magnetic moments.
The exceptions to this rule are
the works of Alexandrou {\it et.~al.}, where the
dependence of $\Delta$ form factors
on $Q^2$  was
studied systematically for the first time
\cite{Alexandrou07b,Alexandrou08c},
and the recent work of the Adelaide group \cite{Boinepalli09}.

Our  work uses a covariant model \cite{NDelta} based on valence 
quark degrees of freedom to calculate the $\gamma \Delta \to \Delta$ form
factors and compares them to state-of-the-art results. 
We have restricted 
the nucleon and the $\Delta$ wave functions to their S-wave components.
With such a symmetric distribution
only the charge and the magnetic dipole form factors 
are predicted, and the remaining form factors, 
the electric quadrupole and magnetic octupole, 
vanish, since they measure the deviation 
from the symmetric distribution. 
Nevertheless, it is still interesting to verify,
as done in this work, whether a $\Delta$ model
calibrated by the N$\Delta$ transition 
gives a good description 
of the lattice QCD 
extrapolation of the $\Delta$ 
charge and magnetic dipole form factors 
to the physical region, and therefore determines uniquely the
$\Delta$ properties.
We note that our results are true predictions of our model
for the $\Delta$ electromagnetic form factors, 
since the parameters of the constituent quark model were beforehand
fixed by the nucleon and  N$\Delta$ electromagnetic transition
form factors data only \cite{Nucleon,NDelta}.
No further
adjusting of parameters or refitting to the $\Delta$ observables
was done.

Since we did not include D-waves in the $\Delta$ wave function,
the quality of the obtained description indicates  that  there is only a small
percentage (less than 5\%)
of D-wave components in the $\Delta$ wave function,
as found in calculations
of the electromagnetic observables for the
$\rm{N}\Delta$ electromagnetic
transition \cite{Newnucleondelta}.
We note that the contribution of the D-states to the 
$\Delta$ form factors comes predominantly 
from the transitions between the $\Delta$ 
S-state and the $\Delta$ D-states, and that
we have checked
that the correction from the D-states
for the $\Delta$ charge form factor is less than 7\% , and for the
magnetic dipole is less that 14\%, for low $Q^2$
($Q^2<$ 1.0 GeV$^2$) \cite{DeltaDFF}.
With an even more realistic and smaller  D-state admixture \cite{LatticeD}
these numbers decrease even further, to 2\% and
11\% respectively.

Another interesting point to explore is that, although the pion cloud 
is essential to describe the $\gamma N \to \Delta$  transition
reaction form factors, as shown in Ref.~\cite{NDelta}, there is not yet  
direct evidence about the importance of the pion cloud
for the elastic channel ($\gamma \Delta \to \Delta$).
The validity of our approximation of neglecting
the pion cloud is to be judged from 
the deviations of our results from the data.

The paper is organized in the following way:
in Sec.~\ref{secII} we introduce the general definitions
for the $\Delta$ form factors and in Sec.~\ref{secIII}
we present the corresponding analytical expressions
in the covariant spectator formalism used here.
In Sec.~\ref{secIV} we present numerical results and
in Sec.~\ref{secV} we draw conclusions.

\section{$\Delta$ electromagnetic form factors}
\label{secII}

The electromagnetic interaction with a on-mass-shell $\Delta$ isobar, 
for initial momentum $P_-$ and final momentum $P_+$, 
can be parametrized in terms of the current 
\cite{Nozawa90,Pascalutsa07,Vanderhaeghen08}:
\ba
J^\mu&=&- \bar w_\alpha (P_+) 
\left\{ 
\left[
F_1^\ast(Q^2) g^{\alpha \beta }
+ F_3^\ast (Q^2)
\frac{q^\alpha q^\beta}{4M_\Delta^2}  \right] \gamma^\mu
\right. \nonumber \\
& & \left.
+ \left[ F_2^\ast(Q^2) g^{\alpha \beta } +
F_4^\ast (Q^2)   
\frac{ q^\alpha q^\beta}{4M_\Delta^2} \right] 
\frac{i \sigma^{\mu \nu}q_\mu}{2M_\Delta} \right\} w_\beta (P_-).
\label{eqJ0}
\ea
where
$w_\alpha$ is the Rarita-Schwinger 
spin 3/2 state, 
associated with the spin projection 
$s=\pm 1/2,\pm 3/2$, which is not specified in the equation  
for the sake of simplicity.
The functions $F_i^\ast(Q^2)$ 
($i=1,...,4$) are the $\Delta$ form factors.
In particular $F_1^\ast(0) =e_\Delta$, where 
$e_\Delta$ is  the $\Delta$ charge. 

It is more convenient to use the multipole form factor
functions, labeled in terms of the 
multipoles observed in the electromagnetic transitions.
The $\Delta$  form factors can be separated into
the electric charge (E0) and quadrupole (E2) form factors, and  
magnetic dipole (M1) and octupole (M3) form factors, 
defined as \cite{Nozawa90,Pascalutsa07,Vanderhaeghen08,Weber78}:
\ba
G_{E0}(Q^2) &=& 
\left[ F_1^\ast -\tau F_2^\ast
\right] \left( 1+ \frac{2}{3} \tau \right) - 
\frac{1}{3}\left[ F_3^\ast -\tau F_4^\ast
\right] \tau \left( 1+ \tau \right) 
\label{eqGE0}\\
G_{M1}(Q^2) &=& 
\left[
F_1^\ast + F_2^\ast\right]
\left(   
1+ \frac{4}{5} \tau
\right)
-\frac{2}{5}
\left[
F_3^\ast  + F_4^\ast\right] \tau
\left( 1 + \tau \right) \label{eqGM1}
\\
G_{E2}(Q^2) &=&
\left[ F_1^\ast -\tau F_2^\ast
\right] - \frac{1}{2}\left[ F_3^\ast -\tau F_4^\ast
\right] \left( 1+ \tau \right)  \label{eqGE2} \\
 G_{M3}(Q^2) &=&
\left[
F_1^\ast + F_2^\ast\right]  
-\frac{1}{2}
\left[
F_3^\ast  + F_4^\ast\right]
\left( 1 + \tau \right) 
\label{eqGM3}
\ea 
where $\tau = \frac{Q^2}{4 M_\Delta^2}$.
The static magnetic dipole ($\mu_\Delta$), 
electric quadrupole ($Q_\Delta$) and 
magnetic octupole ($O_\Delta$) are defined,
in the $Q^2=0$ limit, as
\ba
& &
\mu_\Delta = \frac{e}{2 M_\Delta} G_{M1}(0) \nonumber \\
& &
Q_\Delta = \frac{e}{M_\Delta^2} G_{E2}(0) \nonumber \\
& &
O_\Delta = \frac{e}{2 M_\Delta^3} G_{M3}(0).
\ea

\section{Form factors in a S-wave model}
\label{secIII}

In this work we consider the
wave function obtained within the covariant spectator theory,
as proposed in Ref.\ \cite{NDelta}:
\be
\Psi_\Delta (P,k) =
- \psi_\Delta(P,k) \left(T \cdot \xi^{1 \ast} \right) 
 \varepsilon_P^\alpha \;w_\alpha(P).
\label{eqDeltaWF}
\ee
In the previous equation $\psi_\Delta$ 
is a scalar function that describes the momentum distribution 
of the quark-diquark system in terms 
of the $\Delta$   
and the diquark moments, $P$ and $k$ respectively;  
$\varepsilon_{P}^\ast$ is the polarization state 
associated with the diquark spin \cite{Gross08b}
(the polarization index was omitted); 
$w_\beta$ is the Rarita-Schwinger state, as before, 
and $T\cdot \xi^{1 \ast}$ is the isospin operator 
that acts on a given Delta isospin state
[isospin states are not explicitly included].
As discussed in Ref.\ \cite{NDelta}, 
Eq.\ (\ref{eqDeltaWF}) describes a quark-diquark 
system with total
angular momentum $J=\frac{3}{2}$ and no orbital momentum 
(S-state only), with 
the diquark on-mass-shell, a defining condition of the covariant spectator theory.
The scalar wave function $\psi_\Delta(P,k)$  can be written as
\be
\psi_\Delta(P,k) = \frac{N}{m_s(\alpha_1+\chi)(\alpha_2+\chi)^2},
\label{eqPsiScalar}
\ee
where $N$ is a normalization constant, and 
$\chi$ is defined as
\be
\chi= \frac{(M_\Delta-m_s)^2-(P-k)^2}{M_\Delta m_s}.
\ee
The variables $\alpha_1$ and $\alpha_2$ represent 
momentum range parameters in $1/(M_\Delta m_s)$ units.
The particular dependence of $\psi_\Delta$ 
in $(P-k)^2$ given by Eq.~(\ref{eqPsiScalar}) 
was taken for a good description of the
$\rm{N}  \Delta$ transition form factors 
(see Ref.~\cite{NDelta})  in particular the 
magnetic dipole form factor, which dominates the
reaction.
The inclusion of the factor $1/m_s$ in the scalar wave function
(\ref{eqPsiScalar})  
implies that the diquark mass scales out 
of the transition currents, leading to form factors 
independent of $m_s$ \cite{Nucleon}.
From the fit to the data, we have obtained
$\alpha_1=0.290$ and $\alpha_2=0.393$ 
(model II of Ref.~\cite{NDelta}).


\subsection{The current}

In our model, the the $\gamma \Delta \rightarrow \Delta$ 
electromagnetic current in the relativistic impulse 
approximation becomes \cite{Gross08b,NDelta}
\be
J^\mu=
3 \sum_\lambda \int_k 
\bar \Psi_\Delta (P_+,k) j_I^\mu \Psi_\Delta (P_-,k),
\label{eqJ1}
\ee
where the quark current is defined as 
\ba
j_I^\mu &=&  \frac{1}{6}(f_{1+} + f_{1-} \tau_3 ) \gamma^\mu +
\frac{1}{2}(f_{2+} + f_{2-} \tau_3 ) 
\frac{i \sigma^{\mu \nu} q_\nu}{2M},
\ea
and $M$ is the nucleon mass. 

The functions $f_{1\pm}(Q^2)$ and $f_{2\pm}(Q^2)$ 
describe respectively the Dirac and Pauli 
quark form factors that parametrize the 
structure of the constituent quarks. 
As in Refs.~\cite{Nucleon,NDelta} we used 
the form inspired 
on the vector meson dominance mechanism:
\ba
\hspace{-.8cm}
& &
f_{1\pm}(Q^2)=  \lambda + (1-\lambda) 
\frac{m_v^2}{m_v^2+ Q^2} +
c_{\pm}\frac{Q^2 M_h^2}{\left(M_h^2+Q^2\right)^2}
\label{eqF1q}
\\
\hspace{-.8cm}
& &
f_{2\pm}(Q^2)= \kappa_\pm 
\left\{ 
d_\pm \frac{m_v^2}{m_v^2+ Q^2} +
(1-d_{\pm}) \frac{Q^2 }{M_h^2+Q^2} \right\}.
\label{eqF2q}
\ea
In the previous equations $\lambda$ was adjusted to
the charge number density in deep inelastic limit \cite{Nucleon},  which gave
$\lambda =1.21$. The variable $m_v$ represents a vector meson
($m_v=m_\rho \approx m_\omega$), $M_h$ is a mass 
off an effective  heavy vector meson that simulates the 
short range hadron structure, 
and $\kappa_+$ ($\kappa_-$) is the isoscalar (isovector) 
quark anomalous magnetic moment. 
The anomalous magnetic moments $\kappa_\pm$ are
related with the quark $u$ and $d$ anomalous magnetic moments 
through  
$e_q \kappa_q= \frac{1}{6} \kappa_+ +\frac{\tau_3}{2} \kappa_-$, 
where $\tau_3$ is the isospin operator and 
$e_q=\frac{1}{6} +  \frac{\tau_3}{2}$ 
the quark charge, for $q=u,d$ \cite{Nucleon}.
The functions $f_{i\pm}(Q^2)$ ($i=1,2$) are normalized 
according to $f_{1\pm}(0)=1$ and $f_{2\pm}(0)=\kappa_\pm$.

In this application  
we consider in particular the parametrization 
derived in Ref.~\cite{Nucleon}. 
The anomalous magnetic moments $\kappa_\pm$,
were adjusted to reproduce 
the nucleon magnetic moments, 
corresponding to 
$\kappa_+=1.639$ and $\kappa_-=1.823$. The effective heavy meson  mass is $M_h=2 M$, 
as in model II of Ref.~\cite{Nucleon}. 
The remaining parameters correspond also to  
the model II of this last work, constrained by  a fit the the elastic 
nucleon form factor data.
The vector meson dominance coefficients 
are respectively $c_+=4.16$, $c_-=1.16$ and  $d_\pm=-0.686$.


Using Eq.~(\ref{eqDeltaWF}) the current in Eq.~(\ref{eqJ1}) reduces to 
\be
J^\mu= \left[
\bar w_\alpha (P_+) A^\mu \Delta^{\alpha \beta}  w_\beta (P_-) \right]
I_\Delta,
\ee
where 
\ba
& &
A^\mu= 3 \sum_i T_i^\dagger j_I^\mu T_i, 
\label{eqA} \\
& &
I_\Delta= \int_k \psi_\Delta (P_+,k) \psi_\Delta (P_-,k), 
\label{eqID}
\ea
and $\Delta^{\alpha \beta}$ is the covariant sum 
in the diquark polarization for the 
case $M_+=M_- = M_\Delta$; 
see Refs.~\cite{Nucleon,Gross08b,NDelta} for details.
Explicitly one has 
\ba
\Delta^{\alpha \beta} &=& -g^{\alpha \beta} - 
\frac{P_+^\alpha P^\beta_-}{M_\Delta^2}  + 
2 \frac{(P_++P_-)^\alpha (P_++ P_-)^\beta}{4M_\Delta^2 + Q^2}. 
\ea

The integral $I_\Delta(Q^2)$ defined by Eq.\  (\ref{eqID}) 
is normalized according to \mbox{$I_\Delta(0)=1$},
such that the charge condition $J^0= e_\Delta$ holds
($e_\Delta$ is the $\Delta$ charge) for 
$Q^2=0$. 
 
Performing the sum in the isospin operators, and 
after the spin algebra by using the properties of the 
$w_\alpha(P)$ states,  we can furthermore write $J^\mu$ as
\ba
J^\mu&=&
- \bar w_\alpha (P_+) 
\left\{
\left[ \tilde e_\Delta \left( g^{\alpha \beta} + 
2 \frac{q^\alpha q^\beta}{4 M_\Delta^2 + Q^2} \right)
 I_\Delta  \right]
\gamma^\mu \right\}
w_\beta(P_-) \nonumber \\
& & -
\bar w_\alpha (P_+)
\left\{ 
\left[ \tilde \kappa_\Delta \left( g^{\alpha \beta} + 
2 \frac{q^\alpha q^\beta}{4 M_\Delta^2 + Q^2} \right)
 I_\Delta \right]
\frac{i \sigma^{\mu \nu}q_\mu}{2M} \right\}
w_\beta(P_-).
\label{eqJ2}
\ea
The isospin dependent functions $\tilde e_\Delta$ 
and $\tilde \kappa_\Delta$ are respectively
\ba
& &\tilde e_\Delta (Q^2)= 
\frac{1}{2} \left[ f_{1+}(Q^2) + f_{1-}(Q^2)  \bar T_3  \right] \\
 & &\tilde \kappa_\Delta (Q^2)= 
\frac{1}{2} \left[ f_{2+}(Q^2) + f_{2-}(Q^2)  \bar T_3  \right],
\ea
and $\bar T_3$ is the isospin-$\frac{3}{2}$ matrix defined as
\be
\overline T_3=
3 \sum_{i} T_i^\dagger \tau_3 T_i=
\left[
\begin{array}{rrrr}
\; 3 & \; 0 & 0 & 0 \cr 
         \; 0 & \;1 & 0 & 0 \cr
         \; 0 & \; 0 &-1 & 0 \cr
         \;0 & \;0 & 0 &-3 \cr 
\end{array}
\right].
\ee
The $T_i$ ($i=x,y,z$) matrices represent the isospin-$\frac{3}{2}$ 
to isospin-$\frac{1}{2}$ transition operators 
\cite{NDelta,Pascalutsa05,Pascalutsa07}.
Note that in the limit $Q^2=0$, $\tilde e_\Delta$ is just the 
$\Delta$ charge 
\be
e_\Delta = \frac{1}{2} \left(1 + \bar T_3 \right). 
\ee

For the numerical evaluation of the form factors we took one of the models presented
in Refs.\ \cite{Nucleon,NDelta} which is not isospin symmetric.
Isospin symmetry is broken through $f_{1+} \ne f_{1-}$ 
and $f_{2+} \ne f_{2-}$.
Isospin symmetry 
would imply  $\kappa_+ = \kappa_- = \kappa$.
In this case one has $\tilde \kappa_\Delta= e_\Delta \kappa$ for $Q^2=0$.
For the simplest models applied in Ref.\ \cite{Nucleon} 
to the nucleon, the isospin symmetry is 
exact for the Dirac quark current ($f_{1+} \equiv f_{1-}$),  
giving an almost zero electrical form factor for the neutron.
If in Ref.~\cite{Nucleon} we would have 
taken $f_{2+} \equiv f_{2-}$, exact isospin 
symmetry would be observed, and $G_{En}$ would vanish as well.
This is why in the present application we used models which 
are not isospin symmetric.


\subsection{Spectator Form Factors}

Comparing the currents (\ref{eqJ0}) and (\ref{eqJ2}) we 
conclude that
\ba
& & F_1^\ast = \tilde e_\Delta I_\Delta \label{eqF1}\\
& & F_3^\ast = \eta
\tilde e_\Delta I_\Delta = \eta F_1^\ast\\
& & F_2^\ast = \frac{M_\Delta}{M} \tilde \kappa_\Delta I_\Delta \\
& & F_4^\ast = \frac{M_\Delta}{M}  \eta
\tilde \kappa_\Delta I_\Delta = \eta F_2^\ast, \label{eqF4}
\ea
where $\eta= \frac{2}{1+ \tau}$.

\begin{figure}[t]
\vspace{1.1cm}
\centerline{
\mbox{
\includegraphics[width=2.9in]{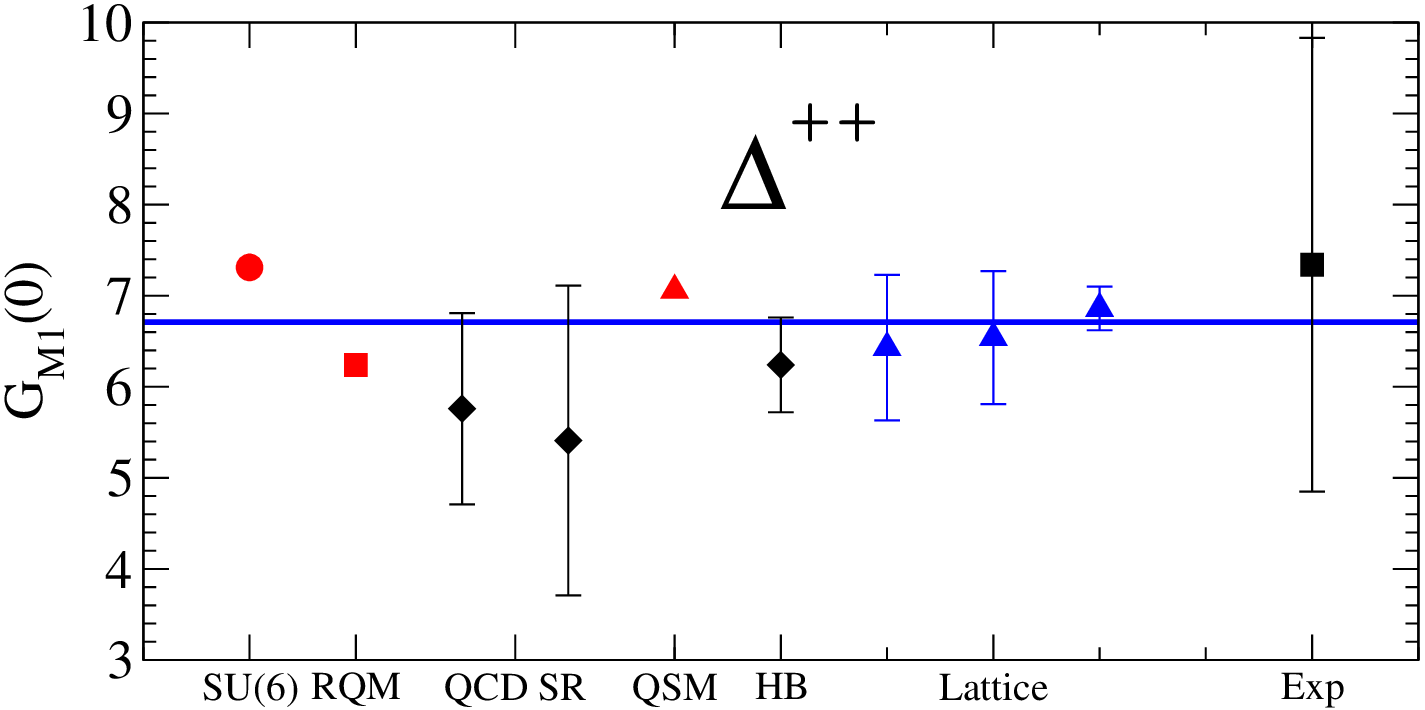}}}
\vspace{1.5cm}
\centerline{\mbox{
\includegraphics[width=2.9in]{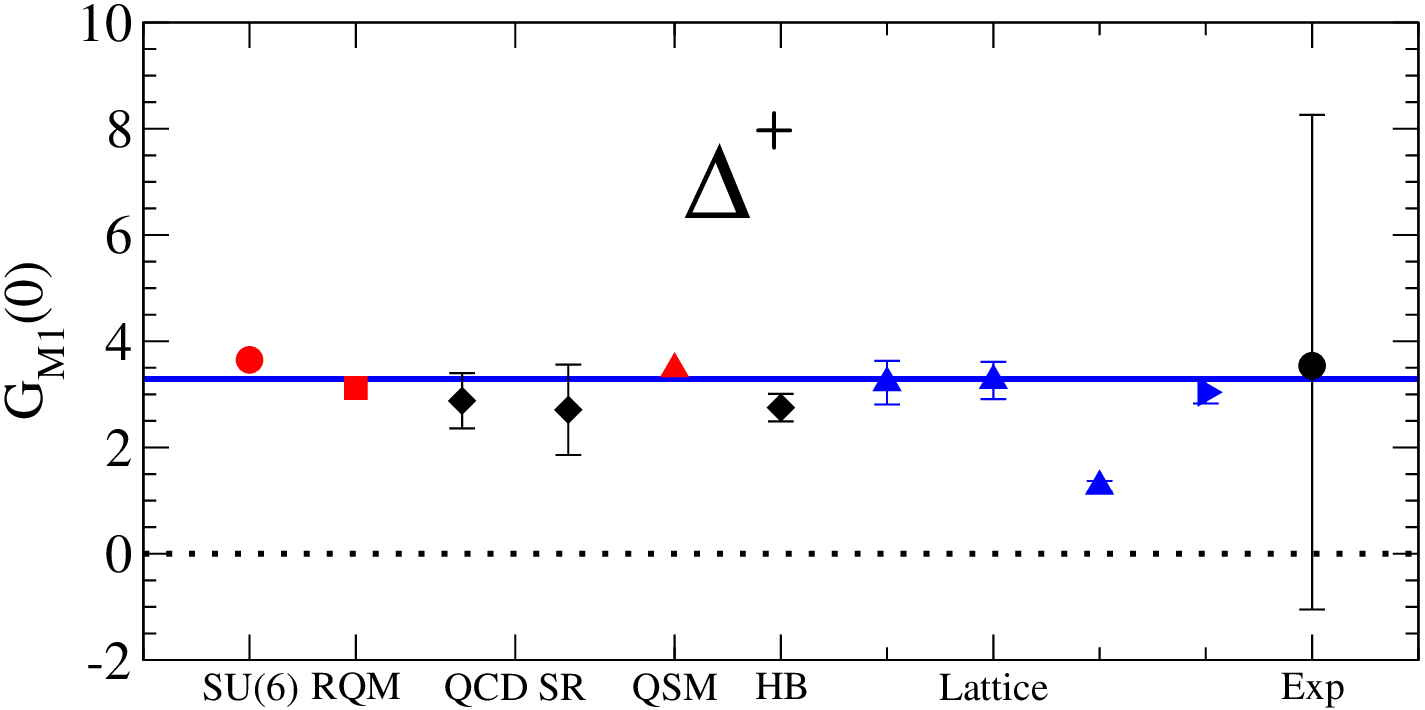}}}
\vspace{1.52cm}
\centerline{\mbox{
\includegraphics[width=2.9in]{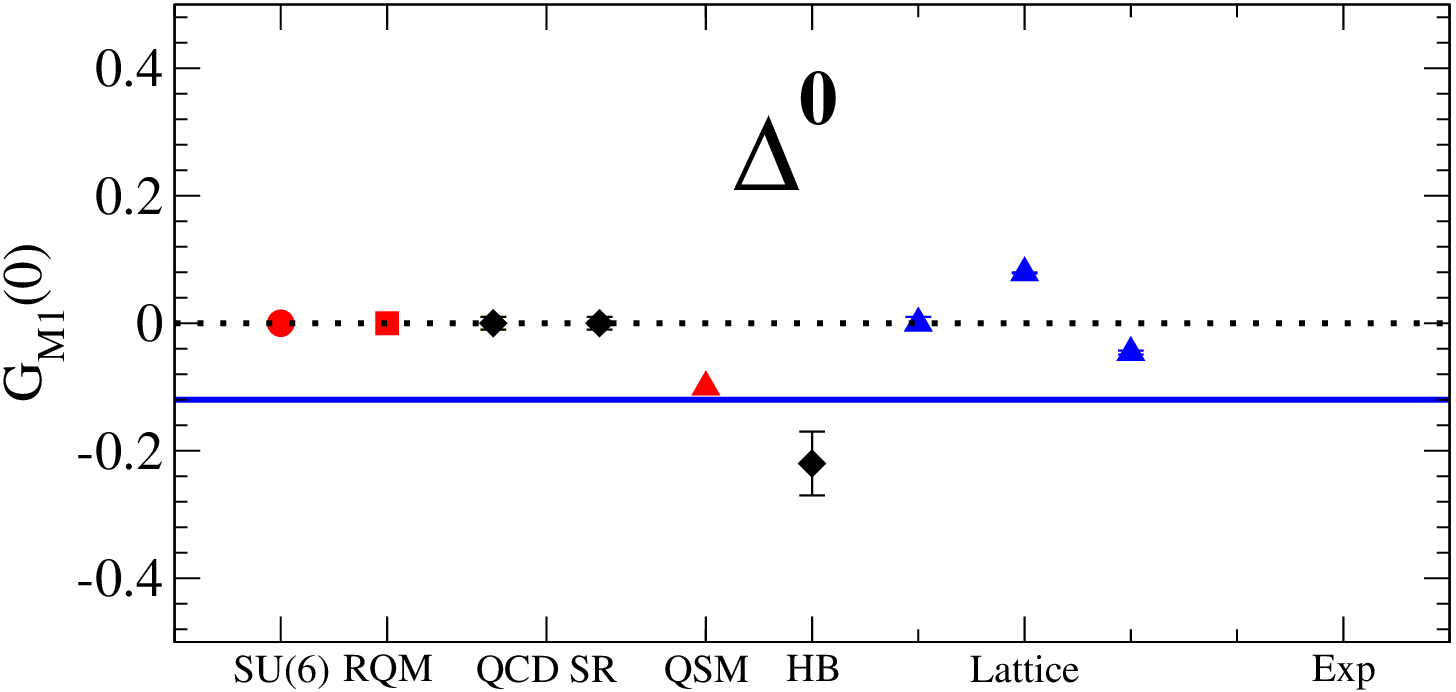}}}
\vspace{1.5cm}
\centerline{\mbox{
\includegraphics[width=2.9in]{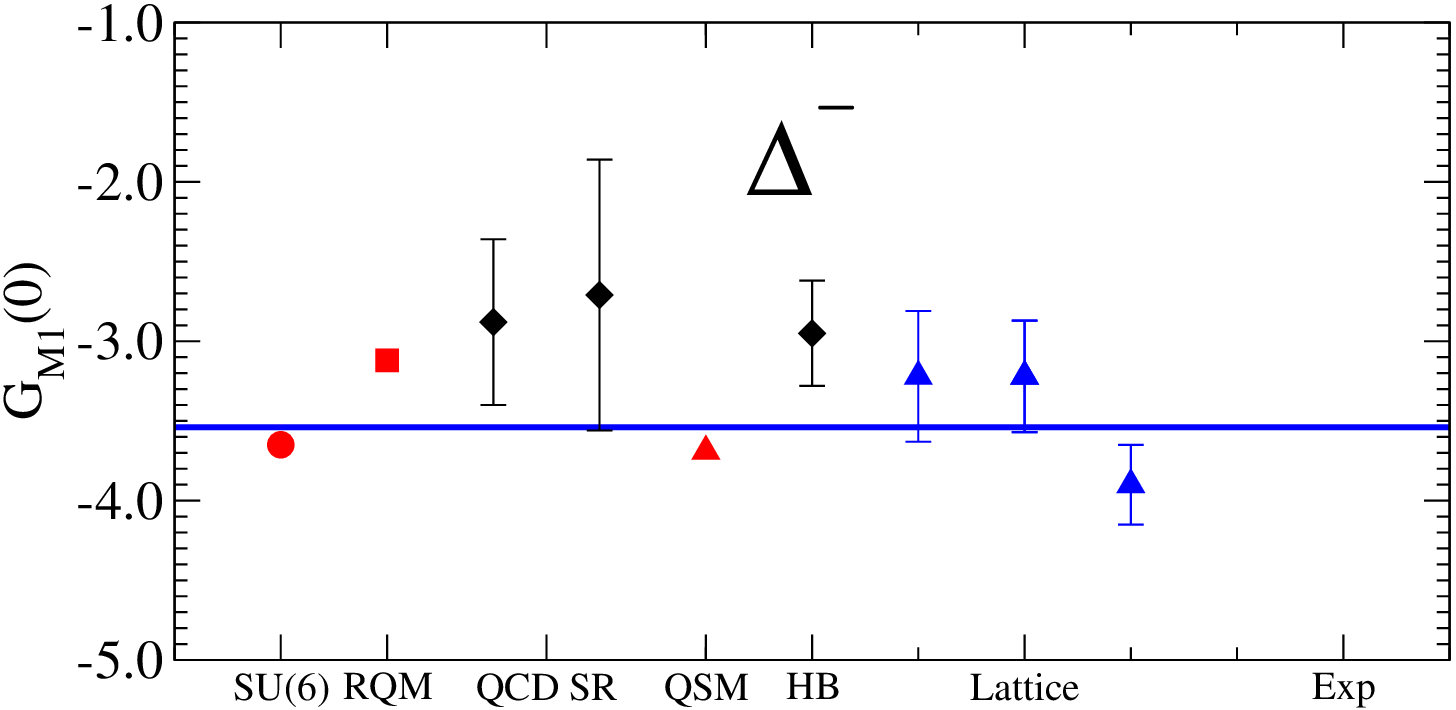}}}
\caption{\footnotesize{$G_{M1}(0)$ in different descriptions.
RQM holds for Relativistic Quark Model,
QSM for ChQSM (Chiral Quark-Soliton Model) and 
HB for HBChPT (Heavy Baryon Chiral Perturbation Theory). 
Lattice results corresponds respectively 
to Refs.\ \cite{Leinweber92,Cloet03,Lee05,Alexandrou07b}.}}
\label{figTable}
\end{figure}

\begin{table*}
\begin{center}
\begin{tabular}{l c  c c c c  c c c}
$G_{M1}(0)$ & & $\Delta^{++}$ & & $\Delta^+$ & & 
$\Delta^0$ & &  $\Delta^-$ \\
\hline
Exp.  & & 7.34 $\pm$ 2.49 & & 3.54$^{+4.59}_{-4.72}$    & & -  & & -   \\
SU(6) & & 7.31            & & 3.65  & & 0  & & -3.65 \\
Rel QM \cite{Schlumpf93}& & 6.24            & & 3.12  & & 0  & & -3.12 \\
QCD SR \cite{Lee98}  & & 5.76 $\pm$ 1.05 & & 2.88 $\pm$ 0.52 & &  0 & &
-2.88 $\pm$ 0.52 \\   
QCD SR \cite{Aliev00}& & 5.41 $\pm$ 1.70 & & 2.71 $\pm$ 0.85 & & 0  & &-2.71 $\pm$ 0.85 \\
ChQSM   \cite{Kim}
 & & 7.06           & &  3.48            & & -0.10 & & -3.69 \\
HBChPT \cite{Butler94}
& &6.24 $\pm$ 0.52 & & 2.75 $\pm$ 0.26 & & -0.22 $\pm$ 0.05  
& & -2.95 $\pm$ 0.33      \\
Lattice \cite{Leinweber92}  & & 6.43 $\pm$ 0.80 & & 3.22 $\pm$ 0.41  & &  0  
& & -3.22 $\pm$ 0.41      \\
Lattice \cite{Cloet03} & &    6.54 $\pm$ 0.73   & &   3.26 $\pm$ 0.35   
& &  0.079  & &  -3.22 $\pm$ 0.35   \\
Lattice \cite{Lee05} & & 6.86 $\pm$ 0.24 & & 1.27 $\pm$ 0.10 & & 
-0.046 $\pm$ 0.003 & & -3.90 $\pm$ 0.25 \\
Lattice \cite{Alexandrou07b} & &           & &    3.04 $\pm$ 0.21 
& &     & &    \\
Spectator  & & 6.71 & &  3.29 & &  -0.12 & &-3.54 \\
\hline
\end{tabular}
\end{center}
\caption{Summary of existing experimental and 
theoretical results for $G_{M1}(0)$.  
The conversion between $\mu_\Delta$ in $\mu_N$ unities to $G_{M1}(0)$ 
is $\frac{M_\Delta}{M} \mu_\Delta$. The Table shows results
based on SU(6) static quark structure 
\cite{Chiang05}, on a relativistic quark model 
(Rel QM) \cite{Schlumpf93}, QCD sum rules (QCD SR)
\cite{Lee98,Aliev00}, Chiral Quark-Soliton Models 
(ChQSM) \cite{Kim}, Heavy Baryon Chiral Perturbation Theory
(HBChPT) \cite{Butler94} and  extrapolations from 
Lattice QCD \cite{Leinweber92,Cloet03,Lee05,Alexandrou07b}. Our results
are labeled "Spectator".}
\label{tableSummary}
\end{table*}

Considering the relations (\ref{eqF1})-(\ref{eqF4}), 
we can write $F_3^\ast-\tau F_4^\ast= \eta ( F_1^\ast-\tau F_2^\ast)$
and $F_3^\ast+ F_4^\ast= \eta ( F_1^\ast +F_2^\ast)$. 
Then, by using the definitions (\ref{eqGE0})-(\ref{eqGM1}), 
we obtain
\ba
& &
G_{E0}(Q^2)= 
\left( \tilde e_\Delta - \tau \frac{M_\Delta}{M} 
 \tilde \kappa_\Delta \right)
 I_\Delta  \nonumber \\
& &
G_{M1}(Q^2)= \left( \tilde e_\Delta + \frac{M_\Delta}{M} 
 \tilde \kappa_\Delta \right) I_\Delta.
\label{eqDeltaFF}
\ea
For the electric quadrupole 
$G_{E2}$ and magnetic octupole $G_{M3}$, 
one has $G_{E2}\equiv 0$ and $G_{M3}\equiv 0$, 
and the electric quadrupole $Q_\Delta$ and
magnetic octupole $O_\Delta$ moments vanish.
This particular result 
follows directly from
our restriction to $L=0$ of the
orbital angular momentum of the 
quark-diquark in the $\Delta$ wave function. 
A better description would require 
other states, in particular
D-state components of the wave function. Still, from 
the N$\Delta$ data, one concludes that the percentage of these
components is small
\cite{Newnucleondelta}.

\section{Numerical results}
\label{secIV}

The model dependence of the $\Delta$ form factors
comes from the quark form factors
$f_{1\pm},f_{2\pm}$ given by Eqs.~(\ref{eqF1q})-(\ref{eqF2q}),
and from the particular form for the
scalar wave function $\psi_\Delta$ given by Eq.~(\ref{eqPsiScalar}),
as introduced in Refs.~\cite{Nucleon,NDelta}.
The quark current is parametrized by three coefficients
($c_+,c_-$ and $d_+=d_-$), and the scalar wave function
by two range parameters ($\alpha_1$ and $\alpha_2$).
Apart  from these five parameters fixed in Refs.~\cite{Nucleon,NDelta},
there are no adjusted parameters involved in the
current calculation.
Then, our results for the $\Delta$ form factors
E0 and M1, given by Eqs.\ (\ref{eqDeltaFF}),
and the magnetic moment $\mu_\Delta$, in particular,
are true predictions.

We evaluate the $\Delta$ electromagnetic form factors numerically by
using Eqs.~(\ref{eqDeltaFF}) for both $Q^2=0$ and $Q^2 \ne 0$.
The first case gives us the static properties of the $\Delta$.
The second case gives us information about the
dynamical properties of the $\Delta$, unfolding
the dependence of the form factors on $Q^2$.

\subsection{Static properties}

The static properties of the 
$\Delta$, like the charge and magnetic moment,
are completely fixed by the static properties 
of the quarks (charge and magnetic moments $\kappa_\pm$), 
constrained by the nucleon charge and magnetic moment.

As mentioned above, for a spherically
symmetric $\Delta$ wave function (S-wave states) 
only the  electrical charge 
$G_{E0}$ and magnetic dipole $G_{M1}$ form factors
are different from zero for $Q^2=0$.
Since the charge of the $\Delta$ is well known, the
relevant information at $Q^2=0$ comes from
the form factor $G_{M1}$ alone. Therefore we will focus
on this observable.

A list of results for $G_{M1}(0)$ obtained from different
frameworks is presented in Table \ref{tableSummary}.
Although additional results can be found in Refs.\ 
\cite{Schlumpf93,Buchmann01,Riska06,Lee98,Aliev00,Lorce08,
Hashimoto08,Machavariani08}, 
the cases 
shown in Table \ref{tableSummary} represent well the
different approaches found in the literature.
The experimental result was taken from the Particle Data Group \cite{Yao06}.
The label "Spectator" denotes model II of Refs.~\cite{Nucleon,NDelta}
used here, and therefore our results.
We present the results for $G_{M1}(0)$ instead of 
the magnetic moment  $\mu_\Delta$ 
in order to simplify the direct comparison with 
the results for the form factor at $Q^2 \ne 0$ in the next section.

All lattice QCD results shown were obtained in the quenched approximation.
In quenched calculations no disconnected diagrams are considered 
\cite{Leinweber92}, and important features of 
quark-antiquark pair creation and pion loop effects 
are suppressed \cite{Young}.
Lattice simulations for heavy pion masses 
are performed in Refs.\ \cite{Leinweber92,Lee05,Alexandrou07b}.
In Refs.\ \cite{Leinweber92,Alexandrou07b} the 
form factors are determined using a linear 
extrapolation of $m_\pi^2$ ($m_\pi$  is the pion mass 
in lattice) to the physical region.
In Ref.\ \cite{Lee05} alternative empirical extrapolations are tested. 
Reference \cite{Cloet03} presents a re-analysis of
the lattice data of \cite{Leinweber92} 
using an effective $\chi$PT.
A similar analysis was also presented in Ref.\ \cite{Pascalutsa05}.
The $\chi$PT effects estimated in Refs.~\cite{Pascalutsa05,Cloet03}
show that the linear extrapolation, 
used in particular in reference \cite{Alexandrou07b} 
for the case $Q^2 \ne 0$, is not totally adequate.
For this reason we 
favor the comparison with lattice 
extrapolation based on $\chi$PT \cite{Cloet03}.
The weak dependence of the data on the 
pion mass suggests nevertheless that a linear extrapolation 
can be taken as good first estimate, meaning 
that the non-analytical terms bring
only small effects,
in contrast with what is required for the
nucleon elastic form factors and 
N$\Delta$ transition form factors.
Unquenched lattice simulations presented 
in Ref.~\cite{Alexandrou08c}, 
and planned extrapolations based on $\chi$PT 
can help to  clarify the situation 
\cite{Alexandrou07b,Alexandrou08c}.

\begin{figure}
\vspace{1.3cm}
\centerline{
\mbox{
\includegraphics[width=3.2in]{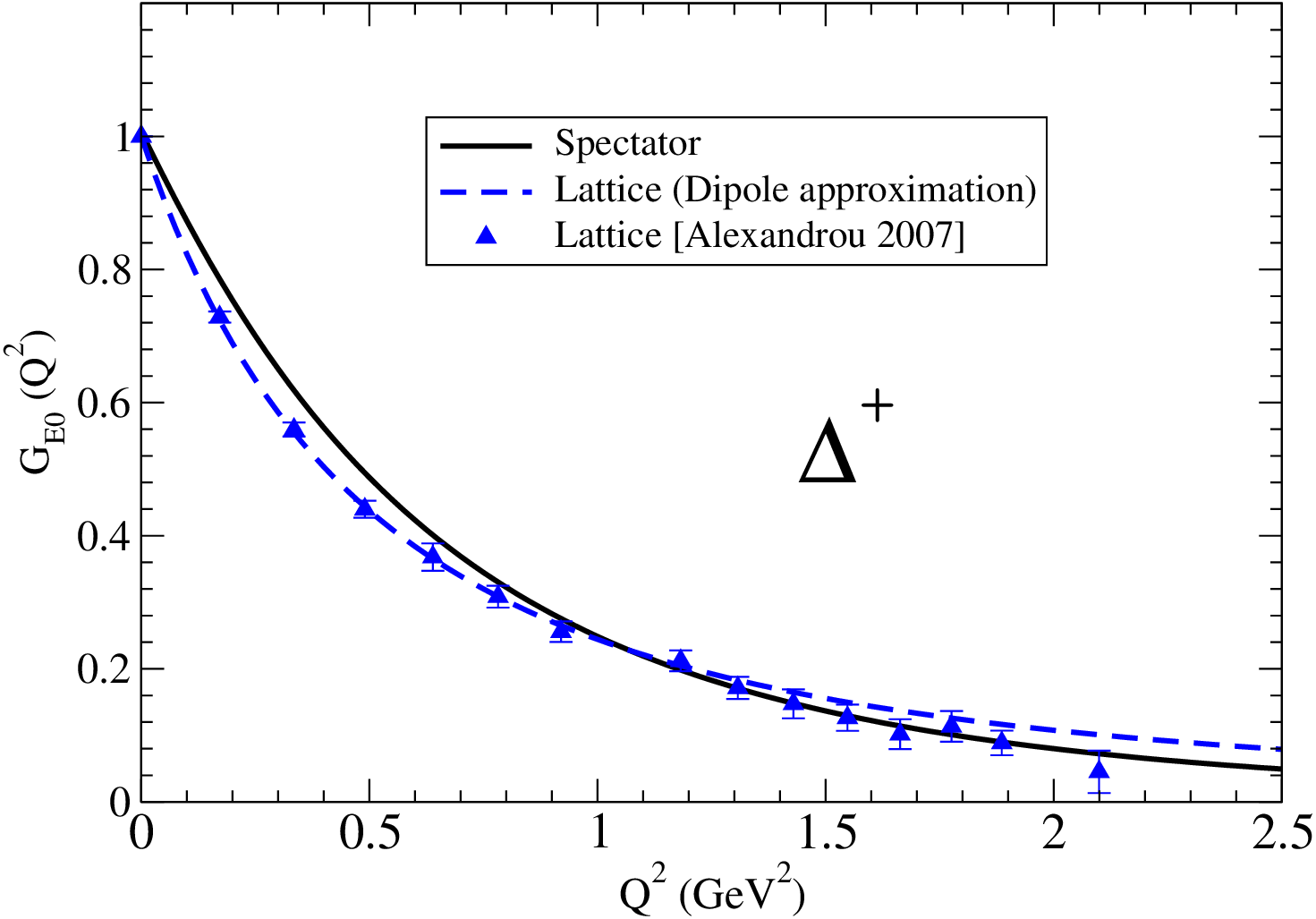} }}
\vspace{0.8cm}
\caption{\footnotesize{Electric charge form factors for the $\Delta^+$.
Lattice data from  
Refs.~\cite{Alexandrou07b,Alexandrou08b}.
The line described as Lattice (dipole approximation)
is a result of a dipole fit from the same references and
corresponds to  $r_{E0}=0.691$ fm.}}
\label{fig1}
\end{figure} 

For a clear comparison of our results with the different 
calculations  we plot in Fig.~\ref{figTable} all 
results, including their uncertainty bars. Fig.~\ref{figTable} depicts
therefore in a graphic form
the results of Table  \ref{tableSummary}.
From the figure we  conclude that,
excluding for now the  $\Delta^0$ state, all
results are consistent with most of 
the predictions within one $\sigma$ deviation.
In particular, for the magnetic moment of
the $\Delta^{++}$, we get a very reasonable value.
Our results are also consistent with 
the central value of the magnetic moment of
the $\Delta^{+}$, although
negative values cannot be excluded due to the error bars.

There is a remarkable exception to the success obtained 
in comparing our results to 
experimental or lattice data for the $\Delta$ baryon: the 
lattice estimation of Ref.\ \cite{Lee05} for  the $\Delta^+$. However,
according to Ref.\ \cite{Young} the lower prediction 
of Lee {\it et.~al.}~\cite{Lee05} is an 
artifact of the quenched calculation 
that, for lower values of the pion mass, 
pushes the magnetic moment of the  $\Delta^+$ aways from 
the proton magnetic moment.

As for the $\Delta^0$ state, it is more difficult 
to extract definitive conclusions. 
From Table I, 
the sign of the $\Delta^0$ magnetic moment is strongly model dependent, and
its magnitude small when compared to the 
other $\Delta$ isospin states. 
Note that there is no direct experimental data,
since the short life time and neutral character of the $\Delta^0$
make difficult the determination of its magnetic moment.
Probably the most significant experimental information 
comes indirectly from measurements on the $\gamma n \to \Delta^0$ reaction.

\subsection{Dynamical properties}

For low $Q^2$ we can approximate any form factor  $G_a$
by a dipole form
\be
G_a(Q^2) = G_a(0)\left( \frac{\Lambda^2}{\Lambda^2+Q^2} \right)^2
\ee
where $a=$E0, M1, E2, M3, 
and $\Lambda$ is an adequate momentum range (cutoff).
In these conditions the low $Q^2$ behavior  
is given by the average 
squared radius $r_a^2$:
\be
r_a^2= \left. - \frac{6}{G_a(0)} \frac{d G_a}{d Q^2} \right|_{Q^2=0}.
\ee

The momentum dependence of the form factors 
was for the first time considered in a systematic way 
in Refs.~\cite{Alexandrou07b,Alexandrou08c}. 
In Ref.~\cite{Alexandrou07b} the $\Delta^+$ state was
considered explicitly, and the lattice data for
the charge form factor $G_{E0}(Q^2)$ 
well simulated by a dipole form factor 
with $r_{E0}=0.691 \pm 0.006$ fm. 
This result is also consistent with the lattice results 
of Ref.\ 
\cite{Cloet03}, $ 0.63 \pm 0.07$ fm.
In Fig.~\ref{fig1} we compare our prediction with 
the lattice data from 
Refs.~\cite{Alexandrou07b,Alexandrou08b}
and also with the dipole form extracted from the data \cite{Alexandrou07b}.
Our prediction does not follow exactly a simple 
dipole form in the range [0,2.5] GeV$^2$,
but nonetheless does not deviate much from the lattice data.
Still,
we predict a slower falloff corresponding 
to  $r_{E0}=0.57$ fm, instead of $r_{E0}= 0.69$ fm.
Therefore, relatively to
the lattice calculations, we predict
an higher concentration of charge at the origin.
Note however, that both lattice calculations and ours
underestimate the proton electric radius, leading to
$r_{Ep}=0.89$ fm \cite{Nucleon}, which implies that 
the $\Delta^+$ has a larger charge density 
near the origin than the proton.
This result contradicts the estimations based on simple
quark models governed by the hyperfine interaction, where
for the $\Delta$
there is a repulsive quark-quark
interaction in the spin-triplet state
which prevents charge concentration at the origin,
differently than for the proton case.
We would then expect a larger extension of the
charge in the $\Delta$ \cite{Giannini90}.
Interestingly, this expectation is contradicted
by the recent quenched lattice QCD simulations \cite{Boinepalli09},
where the $\Delta^+$ charge radius
is always smaller than the proton charge radius.
To clarify this point we need still to wait for unquenched lattice QCD
simulations for pion masses below
the inelastic cut ($m_\pi < M_\Delta-M$).
At any rate, our result is consistent with the speculation
that, contrarily to the $\rm{N} \Delta$ and
 $\gamma N \to N$ transitions
(see models III and IV of Ref.~\cite{Nucleon}),
the pion cloud effect is not as important for the direct reaction
$\gamma \Delta \to \Delta$ form factors, $G_{E0}$ and $G_{M1}$, since
that reaction may be less affected by the behavior of the $\Delta$ as a
$\pi$-nucleon system.
 
Next we discuss the dipole magnetic form factor.
In Ref.\ \cite{Alexandrou07b} only the state $\Delta^+$
was considered, and it was suggested that isospin symmetry 
can be used to obtain the results for the other charge states.
Applying this principle, we consider
\ba
& &
G_{M1}(\Delta^{++}) = 2 G_{M1}(\Delta^{+}) \label{eqD2p}\\
& &
G_{M1}(\Delta^{-}) = -G_{M1}(\Delta^{+}), \label{eqDm}
\ea
in order to generalize the lattice data of Ref.\ \cite{Alexandrou07b}.
Although our model breaks 
the isospin symmetry explicitly ($f_{1+} \ne f_{1-}$ 
and $f_{2+} \ne f_{2-}$), the extent of this violation is very small.
Therefore the comparison of this representation 
with our results is still reasonable
(the exact isospin symmetry gives $G_{M1} (\Delta^0)\equiv 0$).

\begin{figure}[t]
\vspace{1cm}
\centerline{
\mbox{
\includegraphics[width=2.8in]{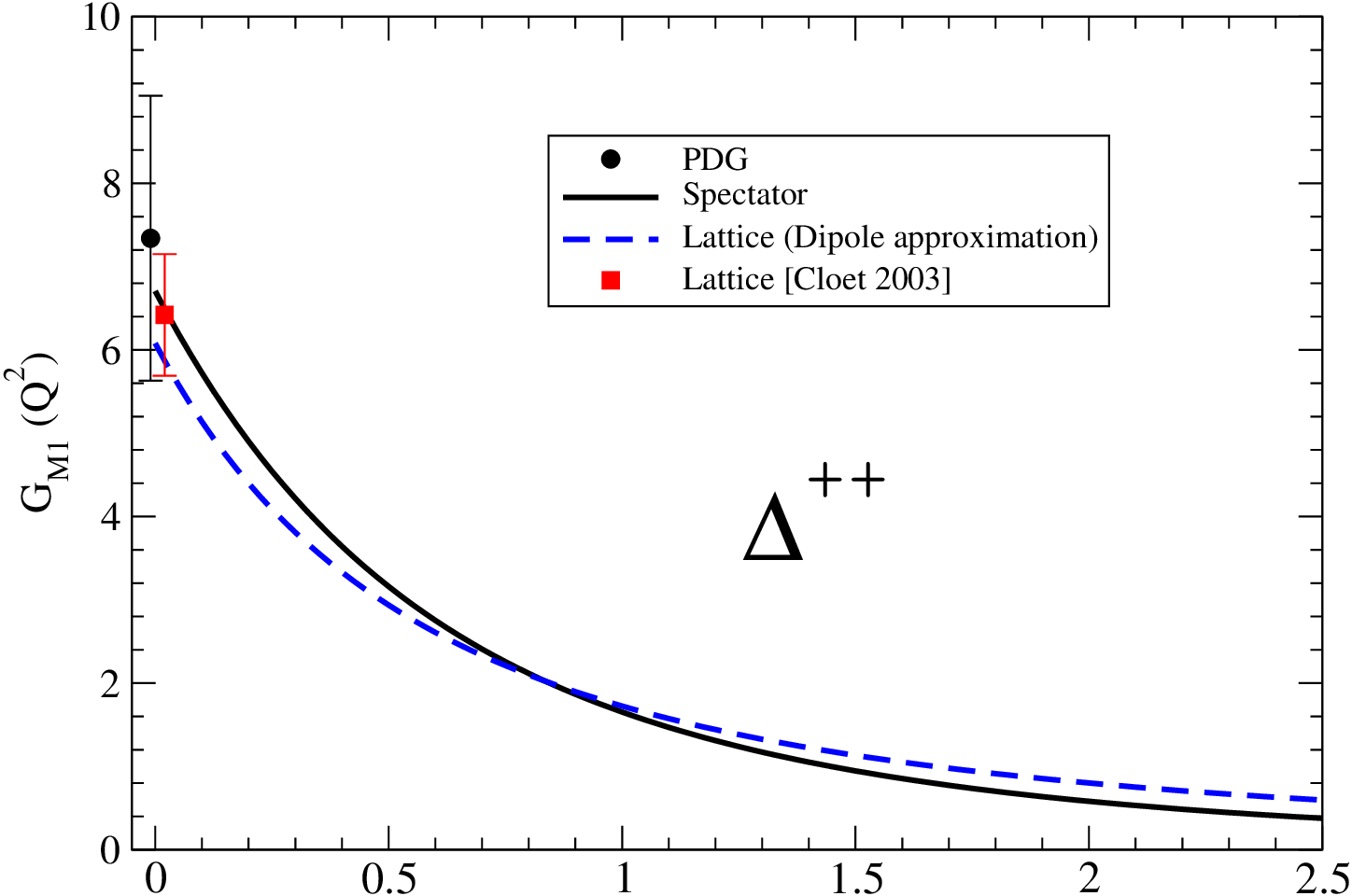} }
\mbox{
\includegraphics[width=2.8in]{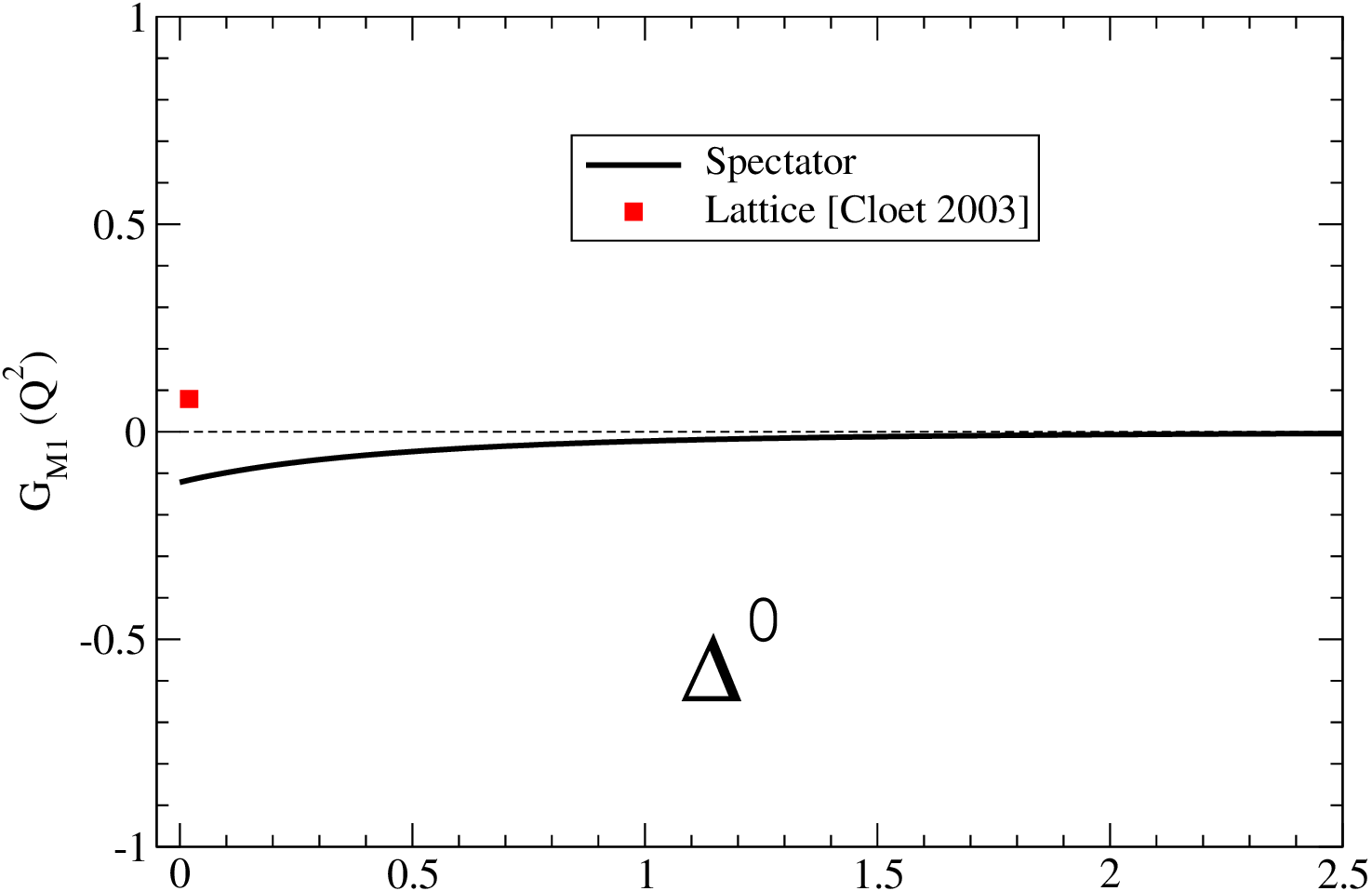} }}
\vspace{1.0cm}
\centerline{
\mbox{
\includegraphics[width=2.8in]{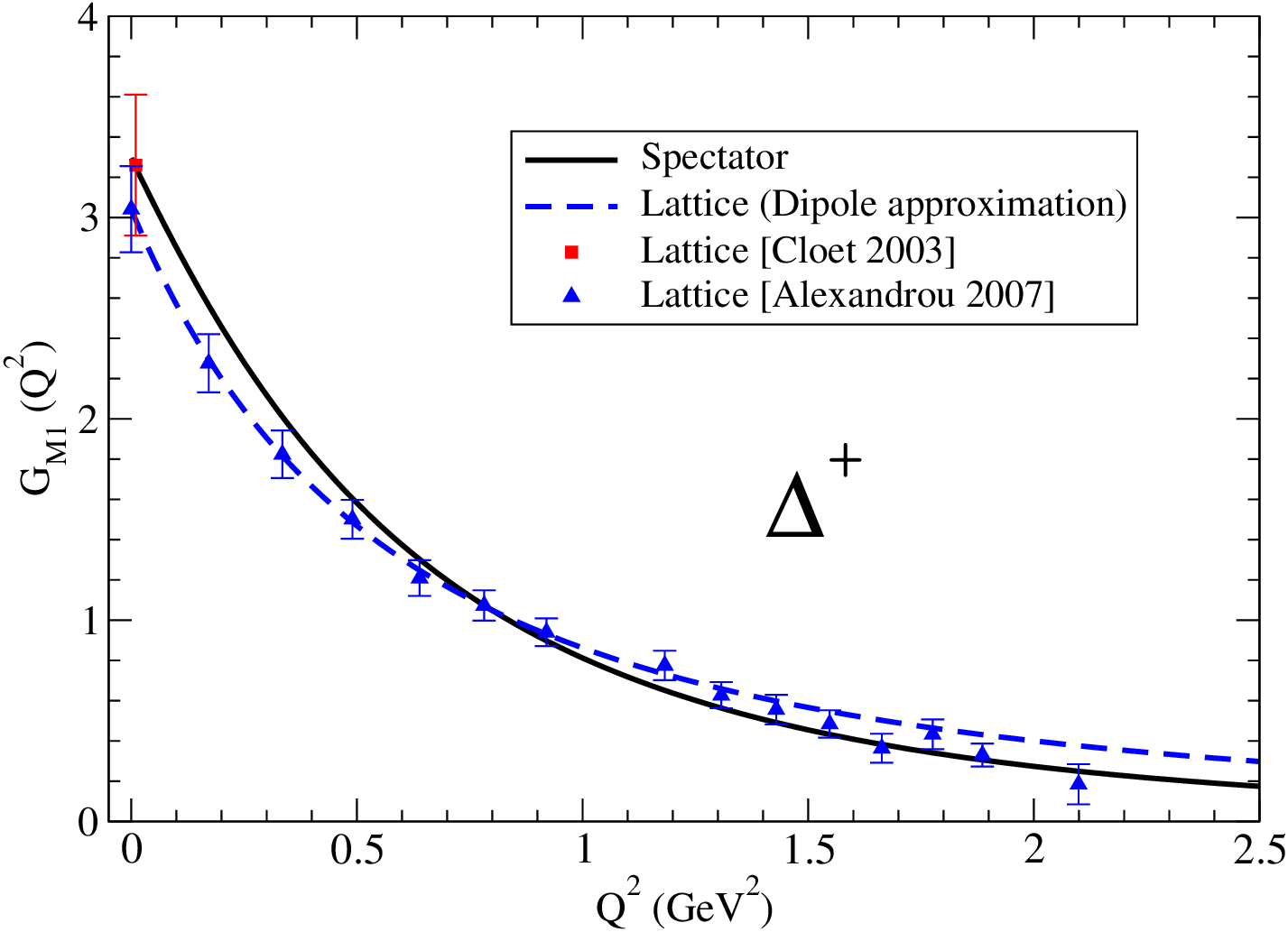} }
\mbox{
\includegraphics[width=2.8in]{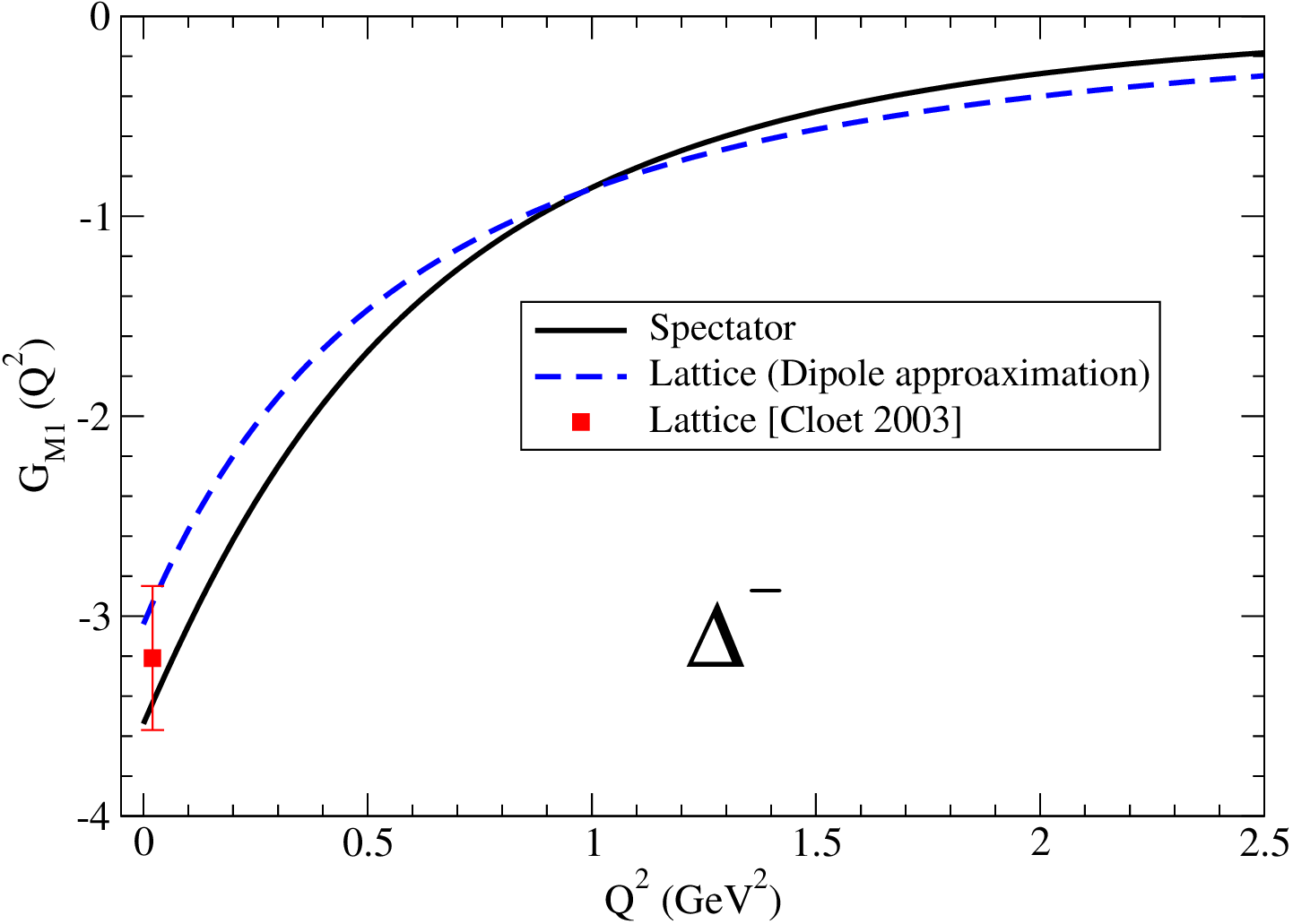} }}
\caption{\footnotesize{
Magnetic dipole form factor for $\Delta^{++}$, $\Delta^+$, 
$\Delta^0$ and $\Delta^-$.
Lattice data from \cite{Cloet03} for $Q^2=0$ and 
Refs.~\cite{Alexandrou07b,Alexandrou08b} for $\Delta^+$.
The dipole line presented for  $\Delta^{++}$ and $\Delta^{-}$
is derived from the lattice data of 
Refs.\ \cite{Alexandrou07b,Alexandrou08b}, assuming 
the isospin symmetry and 
$r_{M1}=(0.642 \pm 0.038)$ fm. 
Experimental data for $\Delta^{++}$
from Ref.~\cite{Yao06}.
The experimental value for $\Delta^+$ is 
not presented (see Fig.~\ref{figTable}). }}
\label{figDelta}
\end{figure}

The results of $G_{M1}$ for each $\Delta$ state are presented in
Fig.~\ref{figDelta}.
We compare our results with 
the chiral extrapolation of the lattice data of Cloet
{\it et.~al.}~\cite{Cloet03},
with the $\Delta^+$ lattice data 
from Refs.~\cite{Alexandrou07b,Alexandrou08b}, 
and with the dipole approximation for the lattice results,
assuming exact isospin 
symmetry as in Eqs.~(\ref{eqD2p})-(\ref{eqDm}), 
using $r_{M1}=(0.642\pm 0.038)$ fm from those works.
For the $\Delta^+$, we estimate $r_{M1}=0.61$ fm 
in agreement with  \cite{Alexandrou07b,Alexandrou08b}.
Our results are 
consistent with the lattice data 
from Alexandrou {\it et.~al.}~\cite{Alexandrou07b,Alexandrou08b}
considering the error bands.
As for $\Delta^-$, 
clearly our results are in agreement
with those of Ref.\  \cite{Cloet03} (for $Q^2=0$)
and are very similar to the dipole approximation.
As for $\Delta^0$ 
no dipole approximation line 
is shown, because exact isospin symmetry 
predicts exactly $G_{M1} \equiv 0$, and 
in fact both our result for 
$G_{M1}(Q^2)$ and the lattice data from \cite{Cloet03} 
are very small when compared with the 
the corresponding results for the charged $\Delta$.
We note that is 
a difference of sign relatively to Ref.~\cite{Cloet03}.
However, this difference of sign is not a problem here, since in
that reference $G_{M1}(0)$ was extrapolated 
from $G_{M1} \equiv 0$ for heavy pion masses,
and no error estimate was given
(see Table \ref{tableSummary}).

\section{Conclusions}
\label{secV}

There is almost no direct experimental information 
about the electromagnetic structure of the $\Delta$ system.
Only for the $\Delta^{++}$ there is some conclusive, although still
very limited data. For this reason 
information on the $\Delta$ structure for $Q^2 \ne 0$ 
has to rely on theoretical models fully constrained by
lattice QCD data, or, inversely, on lattice data extrapolations conveniently
probed in the physical limit.
Several models can be consistent with the data but
different experiments are planned to 
improve the accuracy of the data for the $\Delta^{++}$ and $\Delta^+$, 
which will enable a finer model selection. 
A quark model successful in the description 
of the nucleon elastic form factor data 
and of the dominant contribution to the 
N$\Delta$ electromagnetic transition was taken here to describe the
$\Delta$ form factors, without extra 
parameter fitting \cite{Nucleon,NDelta}.
Our results were compared directly to the (still) scarce existing data
as well as
to the lattice extrapolations to the 
physical region.

In particular, our results compare well
with the first quenched lattice 
study of the $\Delta$ form factors for finite $Q^2$ 
\cite{Alexandrou07b}.
In this work we consider 
only the valence quark degrees of freedom.
Although pion cloud effects are fundamental 
for inelastic reactions, namely the
$\rm{N} \Delta$ transition \cite{NDelta},
our results were obtained without the 
explicit consideration of pion cloud, turning out to be 
consistent with the experimental data
for the $\Delta$ magnetic dipole moment, as well as with 
the extrapolation from lattice QCD \cite{Alexandrou07b} results.
%
As for the charge distribution we 
predict a higher concentration of charge 
near the $\Delta$ center of mass.

There are two aspects to be improved in the future 
for an even more significant comparison. 
On one side, lattice simulations are, until very recently, 
restricted to
quenched approximations 
(Ref.~\cite{Alexandrou08c} is the first unquenched calculation).
Also they depend on extrapolations to 
the physical region that are mostly linear,
without control of the non-analytical terms.
On the other side, our model takes at this stage
a simple spherically symmetric $\Delta$ wave function.
For this reason the subleading form factors 
vanish.
The extension of the model with the 
inclusion of the D-states is being investigated meanwhile \cite{DeltaDFF}.
Upon improvement of both these two aspects in each of the sectors compared
here, the tool developed and tested in this work
will allow one to investigate in finer detail
the failure of the lattice calculations in describing
the form factors for the 
N$\Delta$ electromagnetic transition.

\vspace{0.2cm}
\noindent
{\bf Acknowledgments:}

The authors are grateful 
to Franz Gross for his proposal to initiate the study 
of the baryons within the covariant spectator formalism,
to Constantia Alexandrou for providing us the
lattice data of Refs.~\cite{Alexandrou07b,Alexandrou08b},
to Alfred Stadler for advice during the writing of the text,
and to Marc Vanderhaeghen for having called 
our attention to an error present in a previous 
version. 
G.~R. would like to thank to Ross Young
for helpful discussions.
This work was partially support by Jefferson Science Associates, 
LLC under U.S. DOE Contract No. DE-AC05-06OR23177.
G.~R.\ was supported by the portuguese Funda\c{c}\~ao para 
a Ci\^encia e Tecnologia (FCT) under the grant  
SFRH/BPD/26886/2006.

\end{document}